# An unnoticed side effect of electric vehicles

Tao Wang [a,b], Ying Yang [c], Tieqiao Tang [a] and Xiaobo Qu [a1]

[a]Department of Architecture and Civil Engineering, Chalmers University of Technology, Gothenburg SE-412 96, Sweden

[b]School of Transportation Science and Engineering, Beihang University, Beijing, 100191, China

[c]School of Behavioural and Health Sciences, Australian Catholic University, Sydney 2135, Australia

**Abstract**

We illustrate that the electrification of our transport system might impose unnecessary extra congestion and delay for daily commuting passengers. By modelling travel behaviors of these passengers, it is found that more of them tend to depart at a narrower peak-hour time window. The occurrence of this shift is mainly caused by (1) the energy consumption of electric vehicles (EVs) is much lower than that of traditional vehicles and (2) the energy consumption of EVs is less sensitive to congestion than that of traditional vehicles. We further examine the role of congestion toll in minimizing the extra congestion and delay.

## 1. Introduction

The transport sector contributes approximately 30% of greenhouse gas (GHG) emissions [1,2,3], and that proportion is quickly becoming even higher than that due to rapid urbanization. Transport electrification is well recognized as a panacea for the plight, and there will be increasing societal benefits from electric vehicles (EVs) when the existing energy infrastructure phases out the coal-based power plant, and phases in clean and renewable energy sources, including nuclear energy, biomass energy, and wind energy [4]. EVs do not generate emissions, have a high energy conversion rate, and are able to convert their kinetic energy into electrical energy that can be stored until needed. Thus, the unit travel cost of an EV is far lower than that of a gasoline vehicle (GV). Despite these benefits of EVs, however, they also bring us the problems of a limited range, relatively high fire and explosion risks, and a short lifespan. Nevertheless, it has been widely accepted that the benefits of EVs outweigh their disadvantages. In this regard, many countries, and especially China, are introducing massive incentive policies in an attempt to encourage travelers to purchase and use EVs. China now has the largest EV market and production [5, 6]. According to the International Energy Agency, China will solidify its leadership position in EV sales, with a 28% share, by 2030 [7]. Similar plans are proposed by Norway, the Netherlands, Germany, France, and the UK to ban internal combustion engines by 2050 [8].

---

[1] Corresponding author: Xiaobo Qu, drxiaoboqu@gmail.com, xiaobo@chalmers.se; the first two authors are equally contributed.

When analyzing the problems/concerns regarding EVs, no one has yet considered the link between electrification and congestion. We conduct a simple survey of 37 researchers, practitioners, and government officers involved in EV research and practice. All are found to believe that electrification will impose no substantial impact on traffic congestion, especially during short, morning commuting trips. 26 interviewees mentioned that differences in driving behaviors (caused by the lower capacity for acceleration and deceleration) might result in more congestion, but that this would be marginal for daily commuting passengers. Indeed, it is believed that the current EVs' kinetic performance is good enough for urban road traffic, and the widely cited "range problem" does not play a role in the short trips of the morning commute. The interviewees believed that, for short trips, EVs could be considered perfect substitutes for GVs, with higher efficiency and no emissions.

The above intuitive analysis, however, neglects the fact that the adoption of EVs may change the commuting behaviors of users, especially the time of departure. In this research, we apply a classical travelers' commuting model, which provides the surprising finding that electrification will, in fact, impose very significant delays and congestion during peak hours. Consider a morning commute process, as first proposed by Nobel Laureate Vickery [9], in which most of the travelers make a tradeoff between the cost of congestion, a late arrival penalty, and getting up early (early departure penalty) in their commute. The congestion cost includes travel time, energy cost (petrol or electricity), noise, and emissions [10, 11]. If a traveler replaces a GV with an EV, the energy cost, noise, and emissions will be reduced, and more importantly, they will be less sensitive to congestion due to the regeneration of energy for EVs.

As such, the penetration rate of EVs may result in a new equilibrium among the three cost components: the congestion cost, the early departure penalty, and the late arrival penalty. As the energy cost and emissions become smaller and less sensitive to traffic congestion, it is likely that many EV users will tend to depart during an even smaller time window, which will lead to higher congestion during peak hours if no intervention is employed.

## 2. Results

To explain the EV's influence on traffic congestion, we take a commuting process as an example. In the morning (evening) commute, commuters depart from their homes (workplaces) and travel to their workplaces (homes). If a traveler leaves too early or too late, he/she does not suffer from traffic congestion. However, he/she will suffer from an early departure penalty or a late arrival penalty. Different commuters make distinct selections, and in the long-term, as a result of day-to-day dynamics, the system will reach an equilibrium state. All results in this paper describe the traffic system in the equilibrium state. We illustrate how traffic congestion changes when EVs' market penetration rate (MPR) increases. The results of our investigation are derived from classical commuting behavior models [9, 12], which will be specified in Section 3.

This section shows the impact of EVs' MPR on the traffic congestion in a 20-kilometer trip. The number of commuters during peak hours is 3000 commuters per hour. The commuters who do not suffer from traffic congestion will spend minimal time on the road, which is called the 'free-flow travel time'. In this paper, the congestion delay is defined as the difference between the actual travel time and the free-flow travel time. The congestion delay can be converted into the commuter's average speed, which is often used as the index for congestion [13].

Fig. 1 describes traffic congestion when the EVs' MPR ranges from 0 to 1 in a standard corridor. We define the extra-congested period (ECP) as a performance indicator representing the influence of EVs on the commuting system. The ECP is the time interval in which the commuter's congestion delay is longer than the maximum congestion delay would be if there were no EVs on the road. Fig. 1(A) illustrates that in a situation with purely EVs (MPR = 1), there will be a smaller time window of congestion with substantially higher congestion, in which the maximal on-road delays will increase by up to 80% (traffic volume of 2300 veh/hour vs 3000 veh/hour). Fig. 1(B) reports the performance parameters with respect to different MPRs, ranging from 0 to 1. As can be seen in the figure, the ECP increases with the EVs' MPR, meaning that an increase in the MPR leads to more commuters suffering from severe traffic conditions.

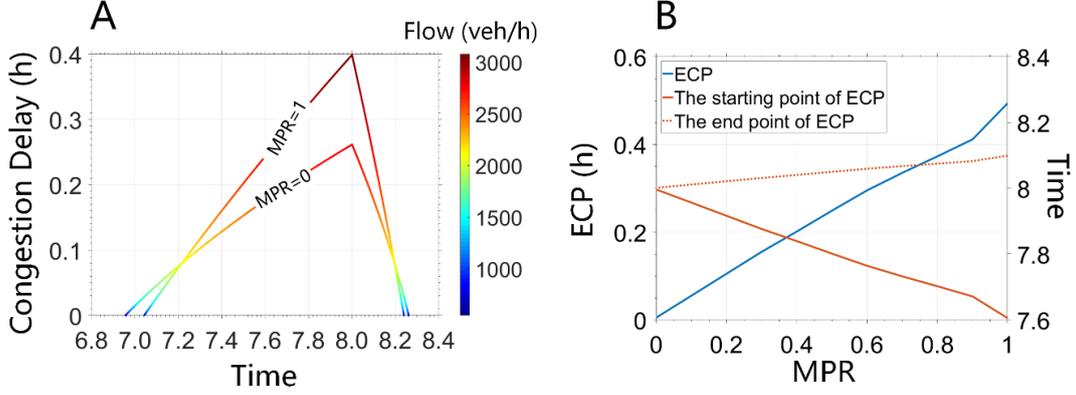

Fig. 1. **Commuters' congestion delays in different cases and the extra-congested period.**

(**A**) EV's MPR = 0 and 1; (**B**) ECP under varying MPR

## 3. Method

### 3.1. Travelers' commuting behavior model

In a commuting process, the commuter's tradeoff between congestion cost, early departure penalty, and late arrival penalty shapes the congestion. Most of the previous literature [9, 12, 15] holds that the individual congestion cost consists only of the travel time cost. However, the cost of energy consumption is actually a substantial element of the personal loss that occurs during the commute. The traveler's travel cost incorporating the vehicle's energy cost is formulated as follows:

$$c(t) = \alpha T(t) + E(T(t)) + \max\{\beta(t^* - t), \gamma(t - t^*)\} \tag{1}$$

where $t$ is the commuter's arrival time at the workplace, $\alpha$ is the unit cost of travel time, $\beta$ is the unit cost of early arrival, $\gamma$ is the unit cost of late arrival, $t^*$ is the preferred arrival time, $T(t)$ is the congestion delay, $E(T(t))$ is the vehicle's energy cost, $\alpha T(t) + E(T(t))$ is the congestion cost, and $\max\{\beta(t^* - t), \gamma(t - t^*)\}$ is the schedule delay cost.

All travelers will choose an appropriate time to arrive at their destination in the equilibrium state. When there is no external interference, the state is called user equilibrium

(UE). After the traffic system reaches UE, no one can improve their own travel time by changing their departure time. There are two steps to deriving the commuter pattern at UE.

**Step 1:** The necessary condition of UE is that all commuters have the same travel cost, i.e., $\frac{\partial c(t)}{\partial t} = 0$, which indicates that Eq. (1) is a family of isocost curves under UE. An example is depicted in Fig. 2.

**Step 2:** We need to determine the equilibrium cost to travelers of selecting a specific trajectory so as to describe how the travel time changes with the system time. In the following, we elaborate on how to obtain the equilibrium cost.

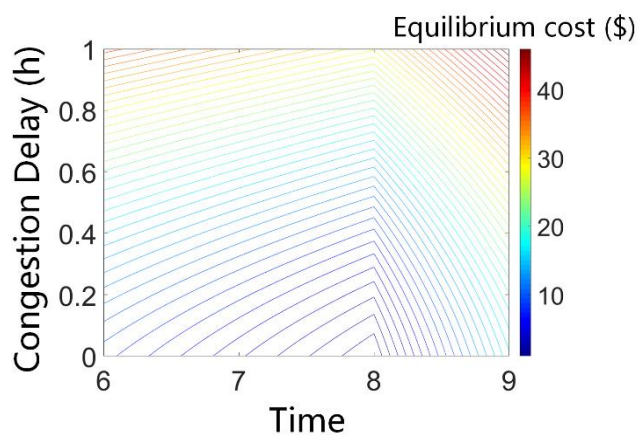

Fig. 2. **A case of the family of isocost curves, in which** $E(T) = 4T + 16.8T^2$, $\alpha = 8.4\ (\$/h)$, $\beta = 4.2\ (\$/h)$, $\gamma = 16.8\ (\$/h)$, **and** $t^* = 8$.

There is a conservation of the number of commuters. This indicates that the time integration of the traffic flow rate at rush hour is the overall number of travelers during rush hour, i.e.,

$$\int_{t_0}^{t_1} f(t)dt = N \qquad (2)$$

where $t_0$ is the starting point of the rush hour of the morning commute, $t_1$ is the ending point of the rush hour, $f(t)$ is the traffic flow rate (the number of vehicles per unit time in a fixed position), and $N$ is the total number of commuters during rush hour.

Almost all traffic flow models show that the travel time is an increasing function of the traffic flow, e.g., see the point queue model [9], the flow speed relation [12, 14], the LWR model [15, 16], and the macroscopic fundamental diagram [17]. The selection of the traffic flow model influences the equilibrium cost for commuters. However, it does not change the shapes of the family of isocost curves. When the total number of commuters in the rush hour and how the flow rate influences the travel time are known, the equilibrium cost can be derived. The travel cost of the first commuter in the rush hour only depends on the arrival time at the destination, i.e., $c(t_0) = \beta(t^* - t_0)$. Then, we can select the specific curve in the family of isocost curves based on the point $(t_0, 0)$.

### 3.2. Energy costs for GVs and EVs

For a fixed-distance trip, the vehicle's energy cost is related to the trip's distance, the travel time of the trip, and some congestion indicators, including the speed variations, the number of full stops, and the coefficient of variation of speed. Smit et al. [18] argued that most of these indicators are decreasing functions (roughly linear functions) of the average velocity. Thus, they increase with travel time. To characterize this non-linear property and make the problem tractable, the energy costs are written as

$$E = \begin{cases} g(T) + g_1 T + g_2 L, & \text{for GV}, \\ e(T) + e_1 T + e_2 L, & \text{for EV}. \end{cases} \quad (3)$$

where $T$ is the travel time of the trip, $L$ is the distance of the trip, $g(T)/e(T)$ is an increasing function of $T$, and $g_2, g_2, e_2, e_3$ are coefficients. The third term, $g_2 L$ or $e_2 L$, is irrelevant to the congestion and is therefore omitted from our formulation.

The functional form of $g(T)/e(T)$ depends on the complexity of the traffic network. For example, in a traffic corridor with fewer ramps or intersections, the traffic flow is usually smoother. Thus, the travel time has little influence on $g(T)/e(T)$, since the long lines at some bottlenecks, other than stop-and-go behaviors, are mostly what make up the delay caused by the congestion. However, in complex traffic networks, greater travel time means a lot of stop-and-go behaviors and a higher coefficient of variation of speed. Therefore, $g(T)/e(T)$ might be an increasing, convex function of the travel time in this case.

Compared to the GV's EC, the EV's EC is less sensitive to the deteriorated traffic condition, due to the EV's high energy conversion rate and energy regeneration process.

Therefore, the EV's EC increases with the travel time at a lower rate than the GV's. As such, the second-order derivative of the EV's EC with respect to the travel time is lower than the GV's as well.

### 3.3. A case of the morning commute

We utilize the flow-speed relation [12, 14] to represent the traffic flow properties of the traffic system as in Eq. (4):

$$\frac{1}{s(t)} = \frac{1}{s_{max}} + \left(\frac{f(t)}{R}\right)^v \quad (4)$$

where $s(t)$ is the average speed when the vehicle arrives at the destination at time t, $s_{max}$ is the free-flow speed, $f(t)$ is the arrival flow rate at time t, $R$ is a parameter related to the road capacity, and $v$ is the elastic variable. In Eq. (3), $T(t) = \frac{m}{s(t)} - \frac{m}{s_{max}}$ ($m$ is the trip distance).

The results of this paper are based on the parameters given in Tables 1. The basic scenario is a commuting process in a complex traffic network, where $g(T)$ is a quadratic function of the congestion delay, i.e., $g(T) = g_2 T^2$. In this case, the EV's EC is $e_1 T + e_2 T^2$.

**Table 1.** Parameters in the basic scenario

| Variable | Value |
|---|---|
| $\alpha$ | 8.4 ($/h) |
| $\beta$ | 4.2 ($/h) |
| $\gamma$ | 16.8 ($/h) |
| $v$ | 4.1 |
| $N$ | 3000 (veh) |
| $R$ | 8000 (veh/h) |
| $m$ | 20 (km) |
| $g_1$ | 4 ($/h) |
| $g_2$ | 16.8 ($/h) |
| $e_1$ | 0.5 ($/h) |
| $e_2$ | 3 ($/h) |

## 4. Mitigation policy –congestion toll design

Our results show that the introduction of EVs imposes substantial extra traffic congestion. Even though EVs reduce the overall energy cost of vehicles and on-road emissions, the imposed congestion in most cases, from the perspectives of the travelers themselves, may outweigh these benefits, especially in the morning commute scenarios. Based on our model, we further propose strategies for balancing the departure times to a larger time window to reduce peak congestion, while maintaining the benefits of lower energy usage and emissions. An intuitive method would be to use time-varying congestion pricing or incentives.

We use a corridor with an MPR of 1 as an illustrative example. Fig. 3 illustrates the optimal design of the time-varying toll (or incentive) scheme. The corresponding performance is also presented. A new indicator, extra congestion delay (ECD), is proposed to explain the benefit of a pricing toll. The ECD is the difference between the travel time when MPR is 1 and the initial travel time when all vehicles are GVs. As can be seen in the figures, the ECD is almost always positive when there is no congestion toll, which indicates that, when all vehicles are replaced by EVs and there is no toll, an overwhelming majority will suffer from severe traffic congestion. However, if the congestion toll is implemented, the congestion will be alleviated substantially.

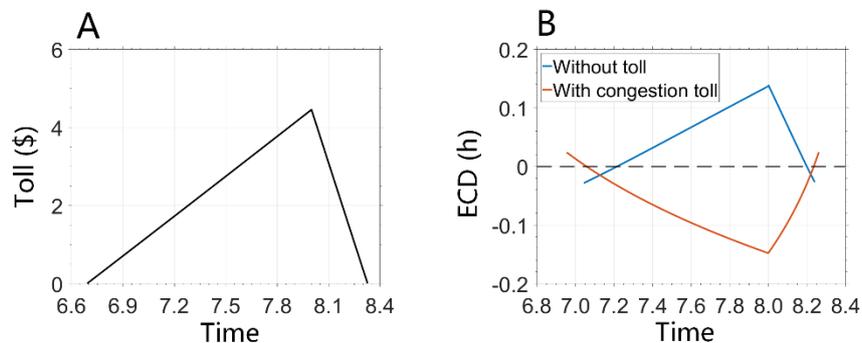

Fig. 3. **The congestion toll and its performance in the case where the EVs' MPR is 1.** (**A**) The optimal congestion toll for the EVs; (**B**) The benefit of the toll to the traffic system.

## 5. Conclusion

Many people believe that electric vehicles (EVs) will soon replace traditional fossil fuel vehicles, as their benefits in terms of energy savings and zero emissions outweigh their demerits such as range limitation and the short life span of the batteries. Most consider EVs to be perfect substitutes for traditional vehicles for short-range trips such as the morning commute. Among the general public, it is believed that the correlation between electrification and congestion is low, especially for short-range trips. In this research, we use a classical morning commuting problem to explain that, without proper intervention, electrification could actually impose additional congestion during the morning peak hours. We derive this finding based on a well-recognized travelers' behavioral model that considers the equilibrium between the early departure penalty, late arrival penalty, cost of travel time, and congestion cost (i.e. energy, noise, and emissions costs). As the energy costs of EVs are much lower than those of traditional vehicles and EVs do not generate emissions, in the above-mentioned equilibrium, the early departure penalty plays a more important role for EV users than for traditional vehicle users. In this regard, in order to avoid the early departure penalty, EV users will tend to travel in a narrower time window during the morning peak hours, which will, in contrast to the abovementioned popular belief, impose extra congestion delays. Based on this observation, we conclude that a transport system comprising EVs might be worse off than one with purely traditional vehicles (as indicated in our example), despite EVs' high energy conversion efficiency and zero emissions. It is therefore critically important to design an intervention strategy in order to reduce the peak congestion while minimizing the total cost. We propose a congestion toll scheme and identify optimal solutions for morning commuting problems. According to our case study, the extra congestion delay caused by the electrification can be totally removed. As such, it is imperative to apply congestion toll for daily commuting trips in the era of electric vehicles.

**Reference**

1.  European Environment Agency (EEA), "Greenhouse gas emissions from transport in Europe" Available at https://www.eea.europa.eu/data-and-maps/indicators/transport-emissions-of-greenhouse-gases/transport-emissions-of-greenhouse-gases-11 (2019). Accessed 23 Feb 2020.

2. Union of Concerned Scientists, "Car Emissions and Global Warming" Available at https://www.ucsusa.org/resources/car-emissions-global-warming (2014). Accessed 23 Feb 2020.

3. Office for National Statistics, "Road transport and air emissions", Available at https://www.ons.gov.uk/economy/environmentalaccounts/articles/roadtransportandairemissions/2019-09-16 (2019). Accessed 23 Feb 2020.

4. Dresselhaus, M.S., Thomas, I.L. Alternative energy technologies. Nature **414(6861)**, 332 (2001).

5. Crabtree, G. The coming electric vehicle transformation. Science **366**, 422-424 (2019).

6. Du, J., Ouyang, D. Progress of Chinese electric vehicles industrialization in 2015: A review. Applied Energy **188**, 529-546 (2017).

7. International Energy Agency, Global EV Outlook 2019 (May 2019) Available at https://www.iea.org/reports/global-ev-outlook-2019 (2019) Accessed 23 Feb 2020.

8. Clewlow, R.R., Mishra, G.S. Disruptive transportation: the adoption, utilization, and impacts of ride-hailing in the United States. Institute of Transportation Studies (2017).

9. Vickrey, W.S. Congestion theory and transport investment. The American Economic Review **59(2)**, 251-260 (1969).

10. Treiber, M., Kesting, A., Thiemann, C. How much does traffic congestion increase fuel consumption and emissions? Applying a fuel consumption model to the NGSIM trajectory data. 87th Annual Meeting of the Transportation Research Board, Washington, DC, 71 (2008).

11. Kaddoura, I., Nagel, K. Simultaneous internalization of traffic congestion and noise exposure costs. Transportation **45(5)**, 1579-1600 (2018).

12. Chu, X. Endogenous trip scheduling: the Henderson approach reformulated and compared with the Vickrey approach. Journal of Urban Economics **37(3)**, 324-343 (1995).

13. Erhardt, G.D., Roy, S., Cooper, D., Sana, B., Chen, M., Castiglione, J. Do transportation network companies decrease or increase congestion? Science advances **5(5)**, eaau2670 (2019).

14. Coria, J., Zhang, X.B. Optimal environmental road pricing and daily commuting patterns. Transportation Research Part B **105**, 297-314 (2017).

15. Newell, G.F. Traffic flow for the morning commute. Transportation Science **22(1)**, 47-58 (1998).

16. DePalma, E., Arnott, R. Morning commute in a single-entry traffic corridor with no late arrival. Transportation Research Part B **46(1)**, 1-29 (2012).

17. Arnott, R. A bathtub model of downtown traffic congestion. Journal of Urban Economics 76, 110-121 (2013).

18. Smit, R., Brown, A.L., Chan, Y.C. Do air pollution emissions and fuel consumption models for roadways include the effects of congestion in the roadway traffic flow? Environmental Modelling & Software **23(10-11)**, 1262-1270 (2008).